\documentclass[12pt]{elsarticle}
\biboptions{sort&compress}
\pdfoutput=1
\usepackage[numbers]{natbib}
\usepackage{url}
\usepackage[hmargin=2.35cm]{geometry}
\usepackage{caption}
\usepackage{subcaption}
\usepackage{xspace}
\usepackage{graphics}
\usepackage{hyperref}
\usepackage{epstopdf}
\usepackage{mciteplus}

\usepackage{ifthen} 
\newboolean{uprightparticles}
\setboolean{uprightparticles}{false} 
\newboolean{articletitles}
\setboolean{articletitles}{true} 

\begin{document}

\title{{\bf Particle Identification at FCC-ee}}

\author[oxford]{Guy Wilkinson}
\address[oxford]{Department of Physics,  University of Oxford, Oxford, United Kingdom} 
%
%
%

\begin{abstract}
Equipping an experiment at FCC-ee with particle identification (PID) capabilities, in particular the ability to distinguish between hadron species, would bring great benefits to the physics programme.  Good PID is essential for precise studies in quark flavour physics, and is also a great asset for many measurements in tau, top and Higgs physics.  The requirements placed by flavour physics and these other applications are surveyed, with an emphasis on the momentum range over which PID is necessary.  Possible solutions are discussed, including classical RICH counters, time-of-flight systems, and d$E$/d$x$ and cluster counting.  Attention is paid to the impact on the global detector design that including PID capabilities would imply.  
\end{abstract}
\maketitle

\section{Introduction}
\label{section:intro}

Particle identification (PID), here defined as the ability to distinguish between hadron species, is essential for several areas of collider physics, in particular flavour and spectroscopy studies, and brings significant benefits for many others.  Until now, conceptual designs for FCC-ee experiments have not placed great emphasis on this capability.  In this brief article the PID requirements at the FCC-ee are assessed, and a summary is given of the various detector technologies that are available to meet these needs.  It is seen that the problem is challenging, and that more studies and developments are  necessary to arrive at a satisfactory solution.

\section{The importance of PID in the FCC-ee physics programme}
\label{section:why}

Among the many studies that will be pursued at FCC-ee, it is flavour physics and spectroscopy, both pursued at the $Z$ pole, that have the greatest reliance on PID.  In flavour physics three main applications exist, which can be illustrated with reference to the decay $B^0_s \to D_s^\pm K^\mp$, which is an important mode in $CP$-violation studies for measuring the Unitarity Triangle angle $\gamma$:
\begin{itemize}
\item{\bf Suppression of same-topology backgrounds} \\ The decay of interest suffers contamination from its sister decay  $B^0_s \to D_s^\pm \pi^\mp$, which is an order-of-magnitude more abundant.  At LHCb this background overlaps with the signal in invariant mass, and must be suppressed by PID.  Although it is expected that the superior mass resolution of FCC-ee detectors will allow the two decays to be separated, this statement is only true if all the decay products are charged.  Any reconstruction involving neutrals, {\it e.g.} $D_s^+ \to \phi\rho^+$, or a study of the equally interesting decay  $B^0_s \to D_s^{\pm \ast} K^\mp$, will only be possible if good $\pi$-$K$ separation is available.  Furthermore, even when studying fully charged final states there will be backgrounds lying under the mass peak from decays such as $B^0_s \to D_s^{\pm \ast} \pi^\mp$, and other channels with missing particles.  Suppression of these background processes will be essential in order to achieve the level of systematic control that will be mandatory with the very large FCC-ee event yields.

\item{\bf Suppression of combinatorial background} \\ In studying rare decay modes involving kaons or protons it is highly advantageous to be able to reject random combinations involving pions.  The same is true when reconstructing the intermediate state, for example the $D_s^+$ meson in the $B^0_s$ decay under discussion.

\item{\bf Flavour tagging} \\ When measuring $C\!P$ asymmetries that are dependent on decay time it is necessary to flavour-tag the event, {\it i.e.} in this case determine whether the decaying $b$-hadron was a $B^0_s$ or $\bar{B}^0_s$ meson at birth.  One powerful tagging method is to inspect the charge of `opposite-side kaons', which arise from the decay chain $b \to c \to s$ ($\bar{b} \to \bar{c} \to \bar{s}$), where the $b$ ($\bar{b}$) indicates the quark (antiquark) produced in association with the beauty antiquark (quark) in the $B^0_s$ ($\bar{B}_s^0$) meson of interest.  In  $B^0_s$ studies it is also possible to look for `same-side kaons', produced in association with the $B^0_s$ meson in the hadronisation process.   PID is essential to exploit both these sources of flavour tag.

\end{itemize}

\begin{figure}[htb]
    \centering
    \resizebox{0.99\textwidth}{!}{\includegraphics{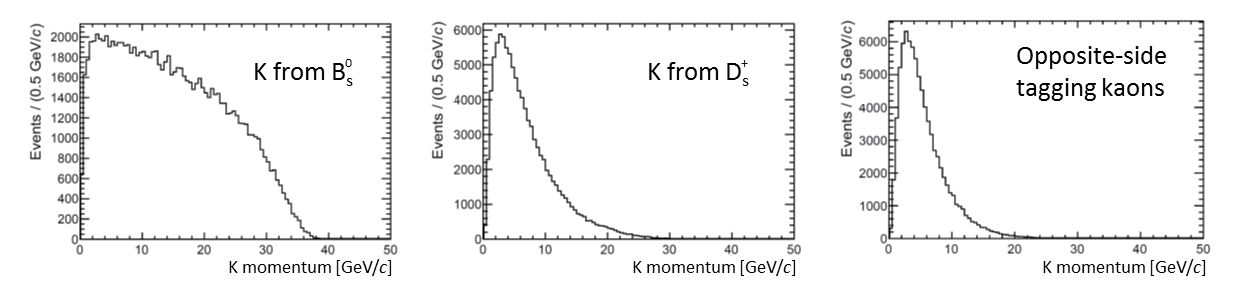}}
    \caption{Momentum spectra for kaons occurring in $Z^0$ events containing a $B_s^0 \to D_s^{\pm}K^{\mp}$  decay.}
    \label{fig:pspectra}
\end{figure}

The momentum spectra of the signal corresponding to these three categories of kaons are shown in Fig.~\ref{fig:pspectra}. The spectrum of the kaons directly from the $B^0_s$ decay is hard, extending up to 40\,GeV/$c$, with around three-quarters of decays contained below 25\,GeV/$c$.  The other two spectra are much softer, with the vast majority of tagging kaons having momentum below 10\,GeV/$c$.  Hence, it is concluded that the PID capabilities at FCC-ee must cover a wide momentum range.

\begin{figure}[htb]
    \centering
    \resizebox{0.42\textwidth}{!}{\includegraphics{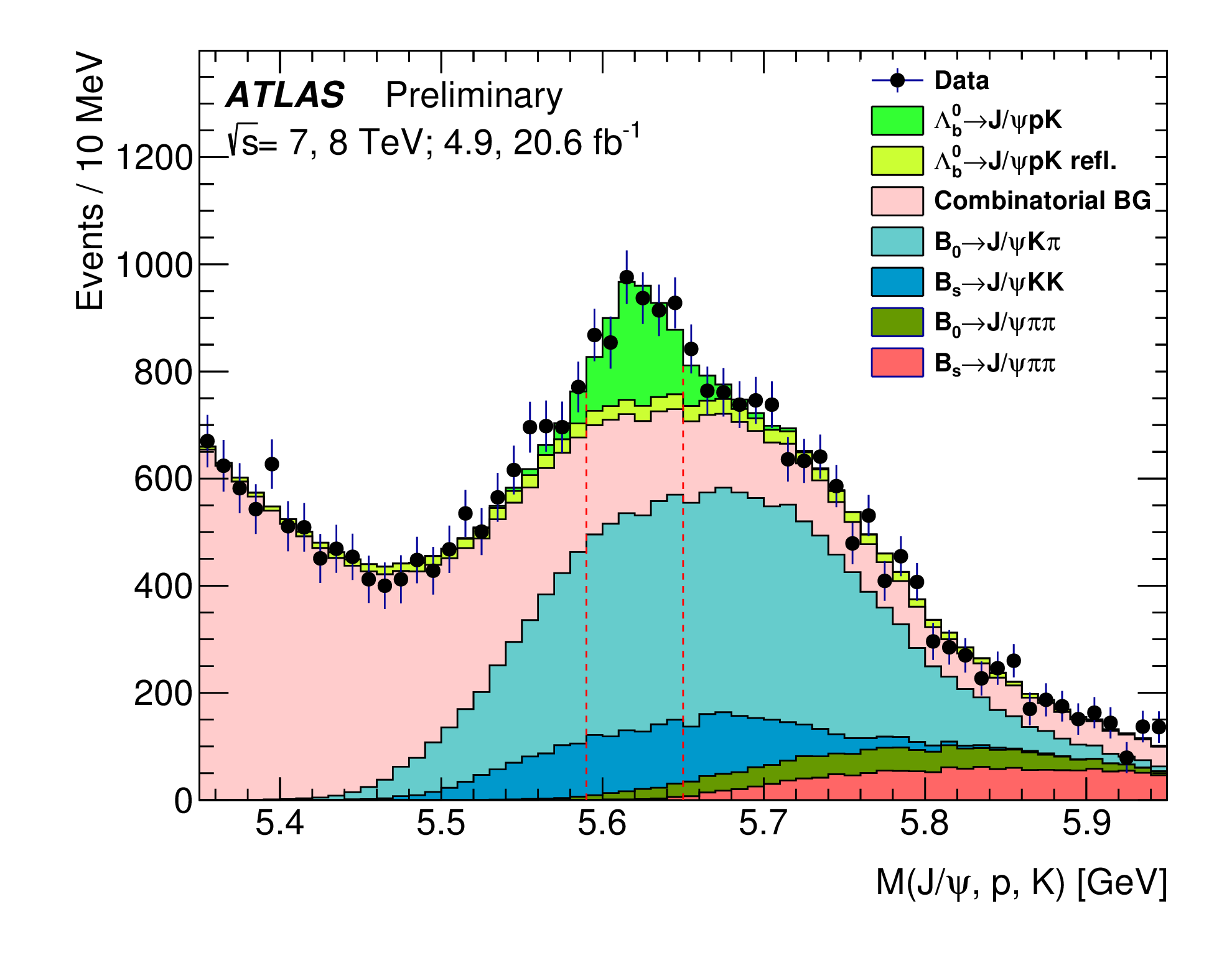}}
    \resizebox{0.48\textwidth}{!}{\includegraphics{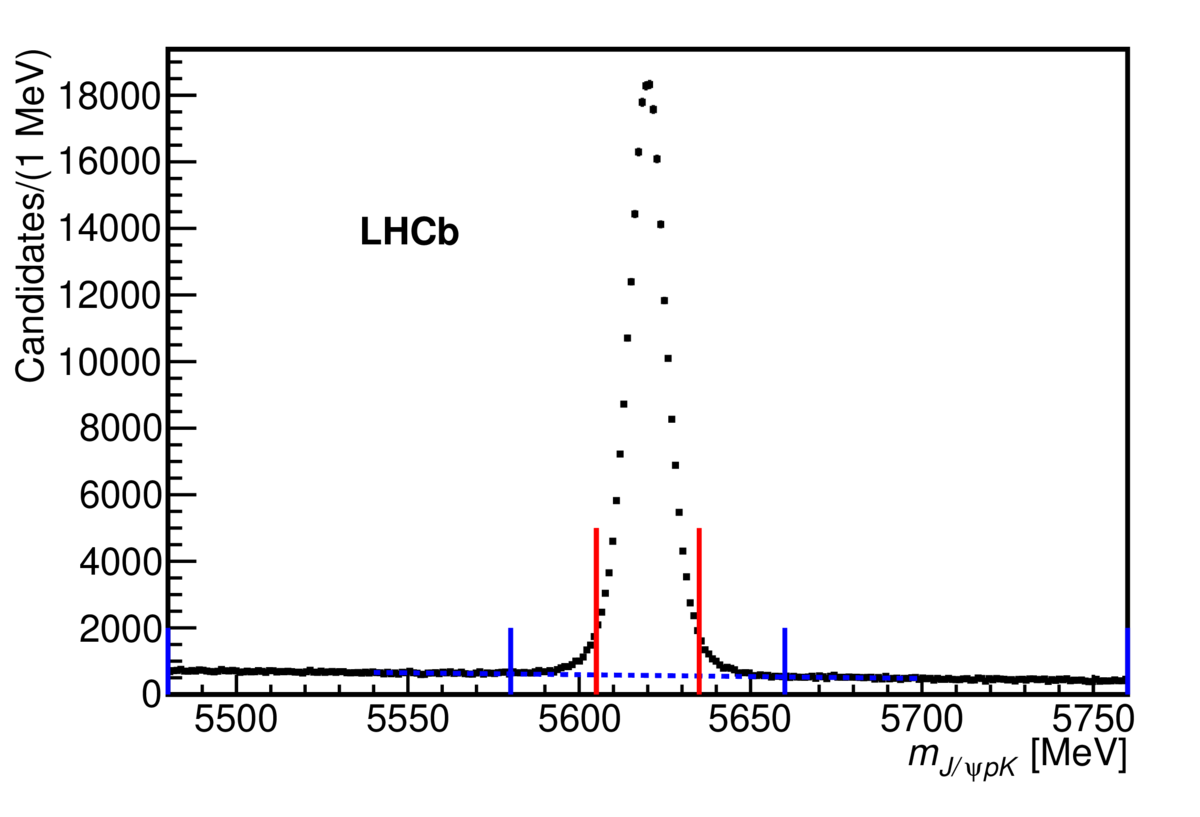}}
    \caption{Invariant-mass spectra of reconstructed $\Lambda^0_b \to J/\psi p K^- $ candidates used in pentaquark studies for (left) ATLAS~\cite{ATLAS-CONF-2019-048} and (right) LHCb~\cite{Aaij:2019vzc} (the vertical lines in the right plot indicate selection windows imposed in the analysis). }
    \label{fig:lambdab}
\end{figure}

Many spectroscopy studies also require the identification of charged kaons and protons.  The benefits of PID can be clearly seen in Fig.~\ref{fig:lambdab}, which shows the invariant-mass spectra of reconstructed $\Lambda^0_b \to J/\psi p K^- $ candidates used in pentaquark studies.  The sample selected by ATLAS, which possesses no PID capabilities, has a high level of background~\cite{ATLAS-CONF-2019-048}, whereas the RICH system of LHCb allows the contamination to be reduced to around 6\%~\cite{Aaij:2019vzc}.  At FCC-ee it is estimated that the momentum spectrum of the protons and kaons in these decays extends to around 20\,GeV/$c$.

PID will be extremely helpful for other studies.  For example in tau analyses it will be valuable to be able to isolate final states containing kaons.  Determining the quark flavour of jets is important in Higgs, top-quark and $W$-physics.  For example, measuring $V_{cb}$, $V_{cs}$ and $V_{ub}$ with on-shell $W$ decays will be a topic of great interest that will greatly benefit from kaon identification.   Although dedicated studies into these possibilities are only beginning, it is assumed that the PID requirements for these measurements are encompassed by those already surveyed for flavour physics at the Z pole. 

\section{Candidate PID technologies}

\subsection{Ring Imaging Cherenkov (RICH) detectors}

\begin{figure}[htb]
    \centering
    \resizebox{0.9\textwidth}{!}{\includegraphics{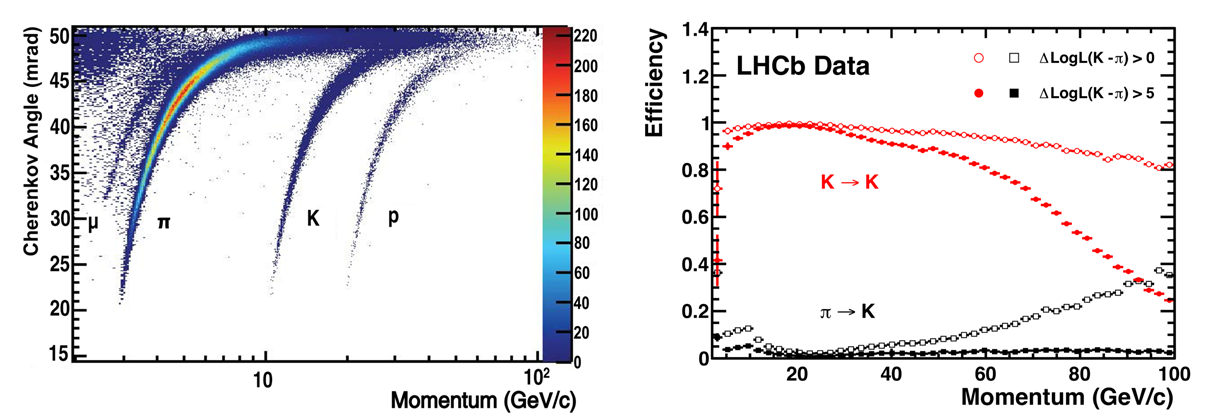}}
    \caption{Performance of the LHCb RICH system.  Left: Cherenkov angle vs. momentum.  Right: kaon identification and misidentication probability vs. momentum, as measured for two working points~\cite{Aaij:2014jba}. }
    \label{fig:LHCbRICH}
\end{figure}

RICHes are the optimum detector technology for providing high-performance PID over a wide momentum range. They were first deployed in collider experiments in the 1990s. The DELPHI RICH at LEP and the SLD CRID at the SLC were dual radiator counters, employing TMAE as a photo-ionising gas~\cite{ALBRECHT199947,CRID}.  The teams who worked on these detectors would be first to acknowledge that they were challenging projects.  PID played an important role in the success of the $B$-factories, but in $\Upsilon(4S)$ decays the required momentum range is low and rather narrow.  For this reason Belle was able to attain satisfactory performance with an aerogel threshold counter, which is not strictly a RICH detector. The DIRC of BaBar was a very elegant and compact solution for this environment, and one which has been further developed in the TOP counter at Belle~II~\cite{Adam:2004fq,Tamponi:2018czd}.  The RICH systems of LHCb and COMPASS have been highly successful~\cite{Adinolfi:2012qfa,Abbon_2007}.  In LHCb, for example, good $\pi$-$K$ separation is available from momenta of a few ${\rm GeV}/c$ up to around 100\,GeV/$c$, as can be seen in Fig.~\ref{fig:LHCbRICH}.  However, these detectors benefit from being allocated a significant amount of space, which allows for an extended radiator length.

Indeed, the limited available space is the greatest challenge when considering the inclusion of a RICH detector in an FCC-ee experiment. With standard gases and current photodetector technologies it is difficult to conceive of a performant layout that has much less than 1\,m of radiator.  The RICH volume must also encompass the mirrors and photodetectors.  If the RICH system is required to provide PID at low momentum, as well as high, then a second radiator must be added, placing further demands on space.  Hence, incorporating a RICH has significant implications for the overall design of the experiment, in particular the size of the tracking volume.  It should also be noted that the gas radiators used in previous and current RICHes are often flurocarbons; environmental considerations demand that other solutions should be sought for future systems.  It is possible that all these problems could be overcome by turning to novel, non-gaseous radiators, such as photonic crystals~\cite{Lin:2018xxy}, but such a technology is at a too embryonic stage to be yet considered as a viable option.

The RICH photodetectors must be efficient for single photons and have good spatial resolution.
If operated within the solenoid of the detector they must be robust against magnetic fields of 2-3~T, although shielding (challenging to accommodate, because of space constraints) and orientation of the devices can weaken this requirement.  
Cost is a major factor, as the total area to be instrumented is large: the total area of the barrel RICH alone would presumably be in the ballpark of $\pi(2\,{\rm m})^2 \times 4\,{\rm m} \sim 50\,{\rm m}^2$ in any detector design, with some reduction factor coming from focusing optics.  
Time resolution, which is an important consideration for other applications, does not appear to be a critical attribute at FCC-ee.
Hybrid Photon Detectors (HPDs), employed successfully at LHCb, and Multianode Photomultiplier Tubes (MaPMTs), which were deployed in HERA-B~\cite{Arino:2003in} and are being installed in the LHCb Run-3 upgrade, have too low a tolerance to magnetic fields to be viable options.   Hybrid Avalanche Photon Detectors (HAPDs) are operating well in the aerogel RICH of Belle-II under a  1.5~T field~\cite{Burmistrov:2020dvn}, but would require further development to be suitable for FCC-ee applications.
Microchannel Plate Photomultiplier Tubes (MCP-PMTs) have shown robustness up to 0.7~T~\cite{HATTAWY201984}, but currently have a high cost per unit area.
Silicon Photomultipliers (SiPMs), although never yet deployed in a RICH sytem, have high robustness against magnetic fields and are a promising and  fast-evolving technology.  Superficially they may appear unsuitable, because of a high dark count which requires cooling down to around $-40^\circ$\,C to ameliorate. However, radiation damage, which accentuates this behaviour, should be less of a concern at an $e^+e^-$ collider, and random noise that is uncorrelated to the ring structure may not be a fatal impediment to pattern recognition in what is a relatively low-occupancy environment. As their name suggests, Large Area Picosecond Photodetectors (LAPPDs) are an attractive candidate solution on account of their coverage~\cite{Adams:2016tfm}, but require further development.  Gaseous detectors with MPGD readout, as used in the COMPASS upgrade~\cite{Agarwala:2018bku}, represent an alternative line of development that is robust against magnetic fields and relatively inexpensive. The CsI photocathode gives these devices sensitivity in the $\lambda < 205\,$nm regime, which has implications for the mirror design.  Here it is also important to take account of ageing effects in the photocathode through ion bombardment, although the environment at FCC-ee should be relatively benign in this respect.

\vspace{0.2cm} 
\noindent {\bf Late remark: }In the final stages of preparing this article an interesting proposal has emerged for a compact RICH system that relies on a pressure of 4~bar to achieve a low gas-radiator length, coupled with aerogel and SiPM readout~\cite{ARC}.  Further studies are required to validity its suitability for FCC-ee.

\subsection{Time-of-flight detectors}

With the typical flight distances available in cylindrical detectors, time-of-flight (TOF) measurements are a suitable method for providing PID in the lowest momentum range of interest.
By way of example, the ALICE TOF detector, based on multigap RPCs, has a resolution of around 80\,ps and provides three sigma $\pi$-$K$ ($p$-$K$) separation up to transverse momenta of 2.5\,GeV/$c$ (4\,GeV/$c$)~\cite{Akindinov:2013tea}.  
The interest in detectors with ${\cal{O}}(10\,{\rm ps})$ time resolution has increased in recent years because of their application in providing pileup mitigation at the HL-LHC. To this end, the HGTD detector of the ATLAS Phase II Upgrade is designed to reach resolutions of around 30\,ps per particle through the use of Low Gain Avalanche Detectors (LGADs)~\cite{CERN-LHCC-2020-007}.  The MIP timing detector of the CMS Phase II Upgrade aims for a similar performance with the same technology in the endcap, and LYSO crystals equipped with SiPMs in the barrel region~\cite{CMS:2667167}. Timing measurements in this regime will also be available in CMS from the new HGCAL endcap calorimeter~\cite{Ochando:2311394}. Similar considerations are driving the design of LHCb Upgrade II, with the VELO, RICH system and TORCH detector (further discussed below) all intended to have precise timing capabilities~\cite{Aaij:2244311}.  Although pileup is not a concern at FCC-ee, all these technologies are potential interest  for use in a PID detector.  Furthermore, precise timing information may have other applications, for example in the search for long-lived particles, or in systematic studies of the characteristics of events from different regions of the interacting bunches.

\begin{figure}[htb]
    \centering
    \resizebox{0.9\textwidth}{!}{\includegraphics{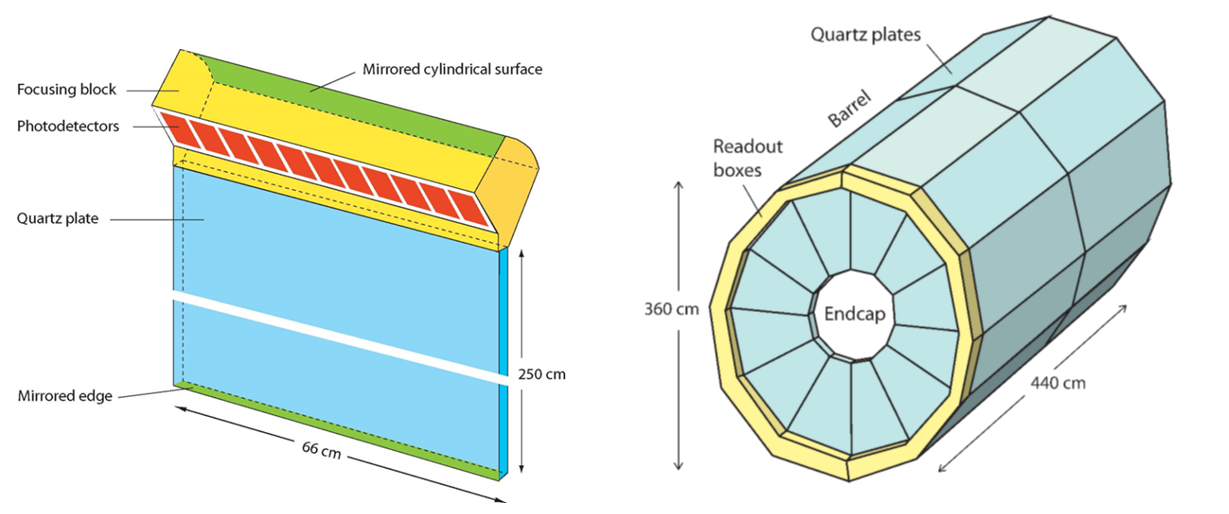}}
    \caption{Left: schematic of a TORCH module, designed for LHCb application.  Right: schematic of a possible TORCH layout in an FCC-ee detector~\cite{RFORTY}. }
    \label{fig:TORCH}
\end{figure}

Cherenkov-based TOF detectors are attracting increasing attention, of which the already mentioned TOP counter of Belle~II is a currently operating example~\cite{Tamponi:2018czd}.  Also worthy of note are the barrel and endcap DIRC detectors of the PANDA experiment at FAIR, which are now under construction~\cite{Schwiening:2019auz}.  All these detectors rely on MCP-PMTs for the photodetectors, which can attain an intrinsic single-channel resolution of better than 10\,ps, although the total single-photon resolution for the whole system is typically an order of magnitude worse.   The most ambitious Cherenkov-based TOF detector is the TORCH (Time Of internally Reflected CHerenkov light)~\cite{CHARLES2011173}.  The TORCH makes a very precise measurement of the Cherenkov angle that allows chromatic dispersion effects to be corrected for, thereby improving the overall time resolution.  Although developed as a standalone detector concept, the TORCH is being considered as a possible component of a future LHCb upgrade~\cite{Aaij:2244311}.  Figure~\ref{fig:TORCH}\,(left) shows a schematic of a TORCH module that would be suited for this application.  Total-internal-reflection of the Cherenkov light occurs within the thin quartz plate, and the radiation is then imaged by MCP-PMTs in a focusing block.  With such a design, and in the LHCb geometry, each photon can be reconstructed with a precision of around 70\,ps, giving around 10-15\,ps when integrating over all the detected photons from a track.  Such a performance would allow for $\pi$-$K$ separation up to around 10\,GeV/$c$.  Beam tests of a prototype module have yielded results that already approach close to the design resolution~\cite{Harnew:2020mnz}.  Thought has been given to how a TORCH system could be integrated into an FCC-ee detector, and a possible layout is shown in Fig.~\ref{fig:TORCH}\,(right).  The thin quartz bars make the concept very attractive from a space point of view,  but a significant volume would still be needed for the focusing block and photodetectors.  As the system would sit closer to the interaction point than in LHCb, improvements in resolution would be required to attain the same level of PID performance.

\subsection{d$E$/d$x$ and cluster counting}

The determination of d$E$/d$x$, the ionisation per unit path length,  is a long-established technique for obtaining PID information in a tracking chamber. Typically this approach is most effective at low momenta ($<$\,$10\,{\rm GeV}/c$).  The measurement is afflicted by Landau fluctuations, which necessitates performing a truncated mean of the different samples accumulated along the track to maintain the discrimination between different particle species.

An alternative method is to count the number of clusters that give rise to the ionisation~\cite{CASCELLA2014127}.  This observable has a Poisson distribution and, in principle, provides better separation.  Exploiting this information, however, requires sufficient resolution and pattern recognition capabilities to identify the individual clusters.  Studies are ongoing to evaluate the feasibility of cluster counting for the TPC of the ILD detector at the ILC~\cite{Behnke:2020krd}.

\begin{figure}[htb]
    \centering
    \resizebox{0.7\textwidth}{!}{\includegraphics{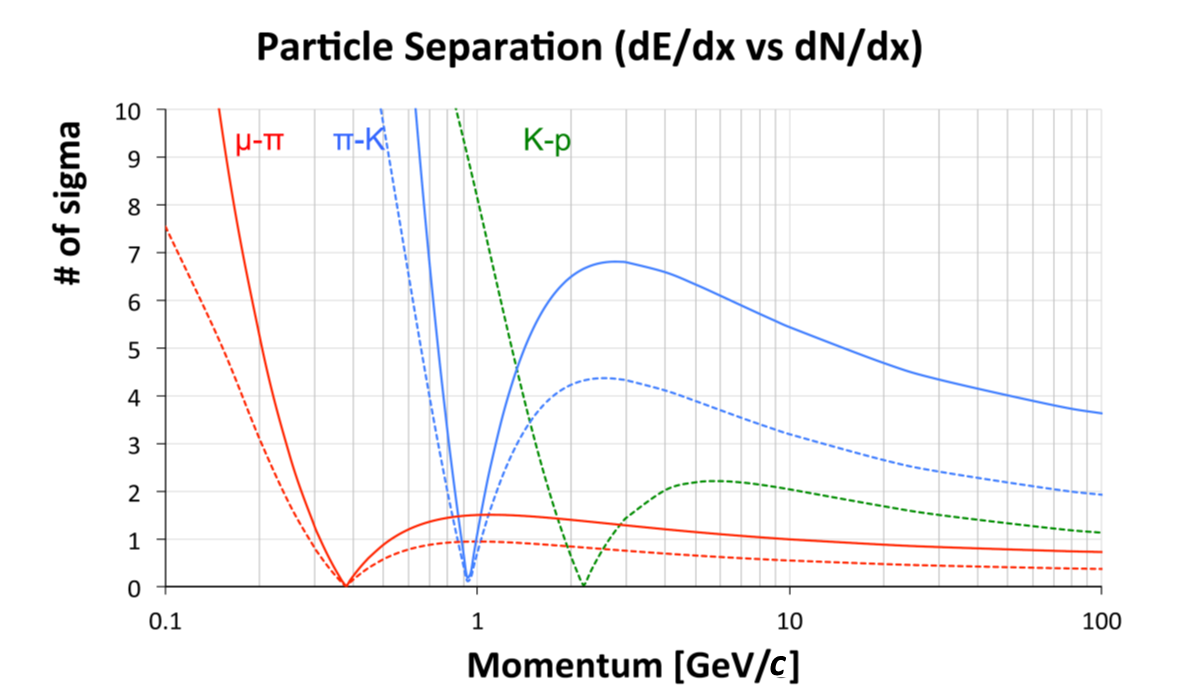}}
    \caption{Expected PID performance of the IDEA drift chamber, both for cluster counting (solid lines) and d$E$/d$x$ measurements (dotted lines)~\cite{Abada:2019zxq}. }
    \label{fig:idea_cluster}
\end{figure}

The drift chamber of the IDEA detector, one of the conceptual experimental designs already advanced for FCC-ee, is intended to have cluster counting capabilities~\cite{Abada:2019zxq}.  The expected PID  performance, expressed in terms of sigma separation between particle hypotheses as a function of momentum, is shown in Figure~\ref{fig:idea_cluster}.  It should be noted that the displayed curves come from an analytic calculation, rather than a full simulation. Nonetheless, the potential advantage of the cluster-counting approach is clear, with a $\pi$-$K$ separation that typically outperforms the d$E$/d$x$ measurement by almost a factor of two at all momenta above a few GeV/$c$. A detector that operated to this specification would, crucially, provide good PID in the higher momentum range of interest.  Both cluster counting and d$E$/dx have a narrow `blind region' around 1\,GeV/$c$, which would need to be compensated for by a complementary PID technology, for example time of flight.

\section{Conclusions}

The enormous physics opportunities at FCC-ee demand that experiments are deployed with a broad range of detector attributes.  Good PID capabilities across a wide momentum range will be essential for flavour and spectroscopy studies, and will enhance the physics reach in many other topics.  The cluster counting that is proposed for the IDEA drift chamber appears a promising solution, but further work is needed to understand the level of performance that can be expected.  Alternative approaches, in particular RICH counters and time-of-flight systems, would be required for other experiments.  The spatial requirements that these detectors impose have implications for the choice and design of the other components of the experiment, and mandate a global optimisation.  Further developments in detector technology are encouraged in order to soften these constraints.

%
%

%
\bibliographystyle{jhep}
\bibliography{references}

\providecommand{\href}[2]{#2}\begingroup\raggedright\begin{thebibliography}{10}

\bibitem{ATLAS-CONF-2019-048}
{\scshape ATLAS} collaboration, \emph{{Study of $J/\psi p$ resonances in the
  $\Lambda^0_b \to J / \psi p K^-$ decays in $pp$ collisions at $\sqrt{s}$=7
  and 8 TeV with the ATLAS detector}},  Tech. Rep. ATLAS-CONF-2019-048, CERN,
  Geneva, Oct, 2019.

\bibitem{Aaij:2019vzc}
{\scshape LHCb} collaboration, R.~Aaij et~al., \emph{{Observation of a narrow
  pentaquark state, $P_c(4312)^+$, and of two-peak structure of the
  $P_c(4450)^+$}},
  \href{http://dx.doi.org/10.1103/PhysRevLett.122.222001}{\emph{Phys. Rev.
  Lett.} {\bfseries 122} (2019) 222001},
  [\href{https://arxiv.org/abs/1904.03947}{{\ttfamily 1904.03947}}].

\bibitem{ALBRECHT199947}
E.~Albrecht et~al., \emph{{Operation, optimisation, and performance of the
  DELPHI RICH detectors}},
  \href{http://dx.doi.org/https://doi.org/10.1016/S0168-9002(99)00320-4}{\emph%
{Nucl. Instrum. Meth. A} {\bfseries 433} (1999) 47}.

\bibitem{CRID}
K.~Abe et~al., \emph{{Results from the SLD barrel CRID detector}},
  \href{http://dx.doi.org/10.1109/23.322821}{\emph{IEEE Trans Nucl Sci}
  {\bfseries 41} (1994) 862}.

\bibitem{Adam:2004fq}
{\scshape BaBar DIRC} collaboration, I.~Adam et~al., \emph{{The DIRC particle
  identification system for the BaBar experiment}},
  \href{http://dx.doi.org/10.1016/j.nima.2004.08.129}{\emph{Nucl. Instrum.
  Meth. A} {\bfseries 538} (2005) 281}.

\bibitem{Tamponi:2018czd}
U.~Tamponi, \emph{{The TOP counter of Belle II: status and first results}},
  \href{http://dx.doi.org/10.1016/j.nima.2019.05.049}{\emph{Nucl. Instrum.
  Meth. A} {\bfseries 952} (2018) 162208},
  [\href{https://arxiv.org/abs/1811.04532}{{\ttfamily 1811.04532}}].

\bibitem{Adinolfi:2012qfa}
{\scshape LHCb RICH Group} collaboration, M.~Adinolfi et~al.,
  \emph{{Performance of the LHCb RICH detector at the LHC}},
  \href{http://dx.doi.org/10.1140/epjc/s10052-013-2431-9}{\emph{Eur. Phys. J.}
  {\bfseries C73} (2013) 2431},
  [\href{https://arxiv.org/abs/1211.6759}{{\ttfamily 1211.6759}}].

\bibitem{Abbon_2007}
P.~Abbon et~al., \emph{{The COMPASS experiment at CERN}},
  \href{http://dx.doi.org/10.1016/j.nima.2007.03.026}{\emph{Nucl. Instrum.
  Meth. A} {\bfseries 577} (2007) 455}.

\bibitem{Aaij:2014jba}
{\scshape LHCb} collaboration, R.~Aaij et~al., \emph{{LHCb detector
  performance}}, \href{http://dx.doi.org/10.1142/S0217751X15300227}{\emph{Int.
  J. Mod. Phys.} {\bfseries A30} (2015) 1530022},
  [\href{https://arxiv.org/abs/1412.6352}{{\ttfamily 1412.6352}}].

\bibitem{Lin:2018xxy}
X.~Lin et~al., \emph{{Controlling Cherenkov angles with resonance transition
  radiation}}, \href{http://dx.doi.org/10.1038/s41567-018-0138-4}{\emph{Nature
  Phys.} {\bfseries 14} (2018) 816},
  [\href{https://arxiv.org/abs/1804.07118}{{\ttfamily 1804.07118}}].

\bibitem{Arino:2003in}
I.~Arino et~al., \emph{{The HERA-B ring imaging Cherenkov counter}},
  \href{http://dx.doi.org/10.1016/j.nima.2003.08.173}{\emph{Nucl. Instrum.
  Meth. A} {\bfseries 516} (2004) 445},
  [\href{https://arxiv.org/abs/hep-ex/0303012}{{\ttfamily hep-ex/0303012}}].

\bibitem{Burmistrov:2020dvn}
L.~Burmistrov et~al., \emph{{Belle II aerogel RICH detector}},
  \href{http://dx.doi.org/10.1016/j.nima.2019.05.073}{\emph{Nucl. Instrum.
  Meth. A} {\bfseries 958} (2020) 162232}.

\bibitem{HATTAWY201984}
M.~Hattawy, J.~Xie, M.~Chiu, M.~Demarteau, K.~Hafidi, E.~May et~al.,
  \emph{Characteristics of fast timing mcp-pmts in magnetic fields},
  \href{http://dx.doi.org/https://doi.org/10.1016/j.nima.2019.03.045}{\emph{Nu%
cl. Instrum. Meth. A} {\bfseries 929} (2019) 84}.

\bibitem{Adams:2016tfm}
{\scshape LAPPD} collaboration, B.~Adams et~al., \emph{{A brief technical
  history of the Large-Area Picosecond Photodetector (LAPPD) collaboration}},
  \href{https://arxiv.org/abs/1603.01843}{{\ttfamily 1603.01843}}.

\bibitem{Agarwala:2018bku}
J.~Agarwala et~al., \emph{{The Hybrid MPGD-based photon detectors of COMPASS
  RICH-1}}, \href{http://dx.doi.org/10.1016/j.nima.2019.01.058}{\emph{Nucl.
  Instrum. Meth. A} {\bfseries 952} (2020) 161832},
  [\href{https://arxiv.org/abs/1812.06971}{{\ttfamily 1812.06971}}].

\bibitem{ARC}
{R. Forty},
  ``\href{https://indico.cern.ch/event/995850/contributions/4406336/attachment%
s/2274813/3864163/ARC-presentation.pdf }{ARC: a solution for particle
  identification at FCC-ee}.''
\newblock FCC-ee week, CERN, July 2021.

\bibitem{Akindinov:2013tea}
A.~Akindinov et~al., \emph{{Performance of the ALICE Time-Of-Flight detector at
  the LHC}}, \href{http://dx.doi.org/10.1140/epjp/i2013-13044-x}{\emph{Eur.
  Phys. J. Plus} {\bfseries 128} (2013) 44}.

\bibitem{CERN-LHCC-2020-007}
{\scshape ATLAS} collaboration, \emph{{Technical Design Report: A
  High-Granularity Timing Detector for the ATLAS Phase-II Upgrade}},  Tech.
  Rep. CERN-LHCC-2020-007. ATLAS-TDR-031, CERN, Geneva, Jun, 2020.

\bibitem{CMS:2667167}
{\scshape CMS} collaboration, \emph{{A MIP Timing Detector for the CMS Phase-2
  Upgrade}},  Tech. Rep. CERN-LHCC-2019-003. CMS-TDR-020, CERN, Geneva, Mar,
  2019.

\bibitem{Ochando:2311394}
{\scshape CMS} collaboration, C.~Ochando, \emph{{HGCAL: A High-Granularity
  Calorimeter for the endcaps of CMS at HL-LHC}},
  \href{http://dx.doi.org/10.1088/1742-6596/928/1/012025}{\emph{J. Phys. :
  Conf. Ser.} {\bfseries 928} (2017) 012025}.

\bibitem{Aaij:2244311}
{\scshape LHCb} collaboration, R.~Aaij et~al., \emph{{Expression of Interest
  for a Phase-II LHCb Upgrade: opportunities in flavour physics, and beyond, in
  the HL-LHC era}},  Tech. Rep. CERN-LHCC-2017-003, CERN, Geneva, Feb, 2017.

\bibitem{Schwiening:2019auz}
{\scshape PANDA Cherenkov Group} collaboration, E.~Etzelm\"uller et~al.,
  \emph{{The PANDA DIRC detectors}},
  \href{http://dx.doi.org/10.1016/j.nima.2019.01.017}{\emph{Nucl. Instrum.
  Meth. A} {\bfseries 952} (2020) 161790},
  [\href{https://arxiv.org/abs/1901.04283}{{\ttfamily 1901.04283}}].

\bibitem{CHARLES2011173}
M.~Charles and R.~Forty, \emph{{TORCH: time of flight identification with
  Cherenkov radiation}},
  \href{http://dx.doi.org/https://doi.org/10.1016/j.nima.2010.09.021}{\emph{Nu%
cl. Instrum. Meth. A} {\bfseries 639} (2011) 173}.

\bibitem{Harnew:2020mnz}
N.~Harnew et~al., \emph{{Status of the TORCH Project}},
  \href{http://dx.doi.org/10.1088/1748-0221/15/04/C04031}{\emph{JINST}
  {\bfseries 15} (2020) C04031},
  [\href{https://arxiv.org/abs/2003.03373}{{\ttfamily 2003.03373}}].

\bibitem{RFORTY}
{R. Forty}, ``\href{https://indico.cern.ch/event/766859/timetable/ }{TORCH: a
  novel concept for PID}.''
\newblock 11$^{\rm th}$ FCC-ee workshop: theory and experiments, CERN, Jan
  2019.

\bibitem{CASCELLA2014127}
M.~Cascella, F.~Grancagnolo and G.~Tassielli, \emph{Cluster counting/timing
  techniques for drift chambers},
  \href{http://dx.doi.org/doi.org/10.1016/j.nuclphysbps.2014.02.025}{\emph{Nuc%
l. Phys. B, Proc. Suppl.} {\bfseries 248} (2014) 127}.

\bibitem{Behnke:2020krd}
T.~Behnke et~al., \emph{{Recent performance studies of the GEM-based TPC
  Readout (DESY Module)}},  in \emph{{International Workshop on Future Linear
  Colliders}}, 2020.
\newblock \href{https://arxiv.org/abs/2006.08562}{{\ttfamily 2006.08562}}.

\bibitem{Abada:2019zxq}
{\scshape FCC} collaboration, A.~Abada et~al., \emph{{FCC-ee: The Lepton
  Collider}: {Future Circular Collider Conceptual Design Report Volume 2}},
  \href{http://dx.doi.org/10.1140/epjst/e2019-900045-4}{\emph{Eur. Phys. J. ST}
  {\bfseries 228} (2019) 261}.

\end{thebibliography}\endgroup

\end{document}